\documentclass[showpacs,color,colordvi]{revtex4}

\usepackage{graphicx}
\usepackage{dcolumn}
\usepackage{bm}

\newcommand{\bec}[1] {\begin{equation}\label{#1} }
\newcommand{\eec} {\end{equation} }
\newcommand{\beq}{\begin{equation} }
\newcommand{\eeq}{\end{equation}}
\newcommand{\bea}{\begin{eqnarray}}
\newcommand{\eea}{\end{eqnarray}}

\begin{document}


\title{An Effective Membrane Model of the Immunological Synapse}

\author{Subhadip Raychaudhuri$^{(1)}$,
Arup K. Chakraborty$^{(1,2,3)}$, and  Mehran Kardar $^{(4)}$}
\affiliation{$^{(1)}$Department of Chemical Engineering,
$^{(2)}$Department of Chemistry,\\ $^{(3)}$Material Sciences and Physical
Biosciences Divisions, Lawrence Berkeley National Laboratory,
University of California Berkeley, Berkeley CA 94720\\
$^{(4)}$ Department of Physics, Massachusetts Institute of Technology,
Cambridge, MA 02139\\
}

\date{\today}

\begin{abstract}
The immunological synapse is a patterned collection of different types of receptors and ligands 
that forms in the intercellular junction between T Cells and antigen presenting cells (APCs) during
recognition. The synapse is implicated in information transfer between cells, and is characterized 
by different spatial patterns of receptors at different stages in the life cycle of T cells. We obtain 
a minimalist model that captures this experimentally observed phenomenology. A functional RG analysis
provides further insights.  
\end{abstract}


\maketitle

T lymphocytes (T cells) are the orchestrators of the adaptive immune response in complex organisms.
A key event during activation of the immune response is T cell recognition of cells that display 
peptides derived from foreign antigens on their surface \cite{abbas}. 
Recent experiments \cite{monks,grakoui,krummel1,lee,stoll} have vividly 
demonstrated that during this recognition process a highly organized pattern of 
different types of receptors and ligands forms in the intercellular junction between T cells and 
antigen presenting cells.  This recognition motif is several microns in diameter, and since it 
is implicated in information transfer between the cells, it is called the immunological synapse.  
Formation of a synapse is also characteristic of an earlier stage in the life cycle of T cells.  
Immature T cells (thymocytes) are selected in the thymus so that they are not activated by peptides 
derived from the organism itself \cite{abbas}.  In the thymus, thymocytes interact with cells
that display self peptides on their surface.  Thymocytes that bind strongly are 
deleted by apoptosis.  The synapses formed during thymocyte selection 
are distinctly different in character \cite{hailman,richie} from those observed during mature 
T cell activation.  
Understanding the mechanisms via which synapses form under different circumstances and the 
biological purpose of creating different spatial patterns of cell surface receptors are active 
areas of research.  In addition to the biological significance, an understanding of these issues 
may also inspire the creation of synthetic mimics that could carry out biomimetic recognition 
tasks which could be useful in applications such as targeted drug delivery.  

In this letter, starting from a model proposed by Qi et al. \cite{chakraborty1}, we develop a minimalist 
model that captures some of the essential physics of synapse formation when apposing membranes contain 
complementary pairs of receptors and ligands.  The model allows us to calculate a {\it phase diagram} 
which delineates the conditions that lead to a {\it transition} from synaptic patterns characteristic 
of mature T cells to those observed during thymocyte selection.  This phase diagram may serve as a 
guide for the design of synthetic analogs.  

Consider two membranes containing complementary pairs of receptors and ligands. The intramembrane
motion of receptors and ligands is determined either by diffusion or a directed velocity toward the
center of the junction \cite{wulfing1}. Complementary receptors and ligands can bind to each 
other if apposed. Different receptor-ligand complexes have different topographical size \cite{shaw}, 
and hence the rate of association depends
upon local intermembrane separation. The receptor-ligand complexes also dissociate at a prescribed rate.   
For the case where there are only two types of receptors-ligand pairs (TCR-MHCp and 
LFA1-ICAM1 in the case of the T cell immunological synapse), the pertinent equations are:
\bea\label{Pkinetics}
&& \frac{\partial C_{T}}{\partial t} = D_{T} \nabla ^{2} C_{T} - 
             k_{on} (z)C_{T}C_{M} + k_{off} C_{TM} - \vec{\nabla} \cdot (\vec{V} C_{T})
, \nonumber \\
&& \frac{\partial C_{M}}{\partial t} = D_{M} \nabla ^{2} C_{M} -
             k_{on} (z)C_{T}C_{M} + k_{off} C_{TM}, \nonumber \\
&& \frac{\partial C_{TM}}{\partial t} = D_{TM}\nabla\cdot \left [ \nabla  C_{TM}
+ \frac{1}{k_BT}C_{TM} \nabla \frac{\delta F}{\delta C_{TM}}
  \right ] + k_{on} (z)C_{T}C_{M} - k_{off} C_{TM}, \nonumber \\
&& \frac{\partial C_{L}}{\partial t} = D_{L} \nabla ^{2} C_{L} -
             k_{on}' (z)C_{L}C_{I} + k_{off}' C_{LI}, \nonumber \\
&& \frac{\partial C_{I}}{\partial t} = D_{I} \nabla ^{2} C_{I} -
             k_{on}' (z)C_{L}C_{I} + k_{off}' C_{LI}, \nonumber \\
&& \frac{\partial C_{LI}}{\partial t} = D_{LI} \nabla\cdot \left [ \nabla C_{LI}
+ \frac{1}{k_BT} C_{LI} \nabla \frac{\delta F}{\delta C_{LI}}
 \right ] +  k_{on}' (z)C_{L}C_{I} - k_{off}' C_{LI}. \nonumber \\
\eea
Here, $C$ represents the concentration of a given molecule, and $z$ is the
separation of the T cell membrane from the APC membrane.
The abbreviations used for different protein molecules are: TCR $\rightarrow$
T, MHCp $\rightarrow$ M, TCR-MHCp $\rightarrow$ TM, LFA1 $\rightarrow$ L,
ICAM1 $\rightarrow$ I, and  LFA1-ICAM1 $\rightarrow$ LI.
The different species diffuse with the corresponding diffusion coefficients $D$.
$\vec{V}$ is a directed velocity of TCR due to cytoskeletal motion \cite{wulfing1}.
Since small values of $\vec{V}$ do not change the qualitative physics \cite{chakraborty4},
for simplicity we do not consider this further in this letter.
The binding and dissociation chemical reactions are controlled by the 
rate constants $k_{on}(z)$ and $k_{off}$.
The local intermembrane separation, $z$, evolves according 
to a time-dependent Ginzburg-Landau equation,
\beq\label{TDGL}
M^{-1} \frac{\partial z}{\partial t} = - \frac{\delta F}{\delta z} + \zeta,
\eeq
where $M$ is a phenomenological parameter for 
the rate at which membrane shapes respond 
to changes in the free energy,
$F = \int d^2x\left[\frac{\gamma}{2} (\nabla z)^{2}  +
    \frac{\kappa}{2} (\nabla ^{2}z)^{2} +
C_{TM}B_{TM}(z) +
C_{LI} B_{LI}(z)\right]$.
The first two terms are due to costs associated with deforming the membranes:
If one of the membranes is supported and undeformed, $\gamma$ and $\kappa$
are the surface tension and bending rigidity of the free membrane;
for similar membranes they are half the corresponding parameters for each membrane.
$B_{TM}(z)$ and $B_{LI}(z)$ denote respectively the bond energy gained from
forming TM and LI complexes for a specific intermembrane separation $z$.

The time scales associated with large-scale membrane
shape changes can be much slower than those associated with binding kinetics and protein mobility.
In such cases, for a given membrane shape, the local species concentrations are
related by
\bea\label{Rconstant}
\frac{C_{TM}(z)}{C_TC_M}=
\frac{k_{on} (z)}{k_{off} }=\frac{K_{on}}{K_{off}}\exp\left[-\frac{B_{TM}(z)}{k_BT}\right],
\nonumber\\
\frac{C_{LI}(z)}{C_LC_I}=
\frac{k'_{on} (z)}{k'_{off} }=\frac{K'_{on}}{K'_{off}}\exp\left[-\frac{B_{LI}(z)}{k_BT}\right].
\eea
Substituting Eq.~(\ref{Rconstant}) in Eq.~(\ref{TDGL}) 
obtains the following effective dynamical equation for $z$
\bea\label{dyn1}
M^{-1}\frac{\partial z}{\partial t} = && \gamma \nabla ^{2} z - \kappa \nabla ^{4} z
\nonumber \\
- && \frac{K_{on}}{K_{off}}C_TC_M\exp\left(-\frac{B_{TM}(z)}{k_BT}\right)B'_{TM}(z)
\nonumber \\
- && \frac{K'_{on}}{K'_{off}}C_LC_I\exp\left(-\frac{B_{LI}(z)}{k_BT}\right)B'_{LI}(z)+\zeta.
\eea
This corresponds to model A dynamics (Eq.~(\ref{TDGL})) for the order parameter $z$
with an effective free energy functional 
\begin{widetext}
\beq\label{Ftilde}
\tilde{F} =\int d^2x \left[\frac{\gamma}{2} (\nabla z)^{2}  +
    \frac{\kappa}{2}  (\nabla ^{2}z)^{2}
-k_BT\frac{K_{on}}{K_{off}}C_TC_Me^{-\frac{B_{TM}(z)}{k_BT}}
-k_BT\frac{K'_{on}}{K'_{off}}C_LC_Ie^{-\frac{B_{LI}(z)}{k_BT}}
\right].
\eeq
\end{widetext}
The reduction of the full set of equations to one determining the separation $z$ is
reminiscent of the elimination of counter-ions in constructing the
Poisson--Boltzmann equation, where the concentration of counter-ions is obtained
from the Boltzmann weight of a potential, itself determined self-consistently
from the counterion charge profile. Here $z$ plays the role of the potential, while
the protein complexes are analogous to the counterions.

Bound receptor-ligand complexes do not favor membrane shape changes that deform 
the bond from its natural length. We thus use a harmonic approximation for the
functional form of the bond elasticity energy, i.e. 
$B_i(z) =B_i^0+ \lambda_{i} (z-z_{i})^{2}/2$,
with $i=1,2$ for the two possible complexes\cite{chakraborty1}. 
The maximal bond energies $B_i^0$ can be absorbed into $K_{on}/K_{off}$, 
and shall be ignored henceforth.
The natural lengths of receptor-ligand proteins are taken
to be those appropriate for the T cells
($z_{1}=z_{TM} = 15 nm$ and $z_{2}=z_{LI} = 42 nm$) \cite{shaw,garboczi},
with corresponding `spring constants' of $\lambda_i$ given below.
With this approximation, the dynamical Eq.~(4) takes the form (measuring time
in units such that $M=1$)
\beq\label{dyn2}
\frac{\partial z}{\partial t} = - \frac{\delta}{\delta z} 
\left\{\int d^2x\left[\frac{\gamma}{2} (\nabla z)^{2}
+\frac{\kappa}{2}(\nabla^{2}z)^{2}+V(z)\right]\right\}
+ \zeta,
\eeq
with an effective potential
\beq\label{V(z)}
V(z) =k_BT
\left \{ \begin{array}{ll}
-C^*_{1}\exp\left\{ -\frac{(z-z_{1})^2}{2 \sigma_{1}^{2}}
\right \} -
C^*_{2}\exp\left\{ -\frac{(z-z_{2})^2}{2 \sigma_{2}^{2}}\right \},
    \quad \textrm{for $z > 0$}  \\
c, \quad \textrm{for $z < 0$}.
\end{array} \right. 
\eeq
In writing Eq.~(\ref{V(z)}) we have assumed that $\sigma^2 = k_{B}T/\lambda$, which 
is appropriate if the curvature of the receptor-ligand interaction potential is the same 
in the bound state and at the activation barrier. Other assumptions do not alter the
qualitative physics. 
The parameters $C^*_1=({K_{on}}/{K_{off}} )C_{T} C_{M}$ and
$C^*_2=({K'_{on}}/{K'_{off}}) C_{L} C_{I}$ represent the number of complexes 
(TM and LI, respectively)  formed if the intermembrane separation equals the
corresponding optimal bond separation.
As depicted in Fig.~1, the effective potential has two minima corresponding to the
two possible bonds. 
Negative separations are prevented by a large energy cost $k_BTc\gg1$
(hard wall in simulations).

\begin{figure}[b]
\includegraphics[height=3in,width=3in,angle=0]{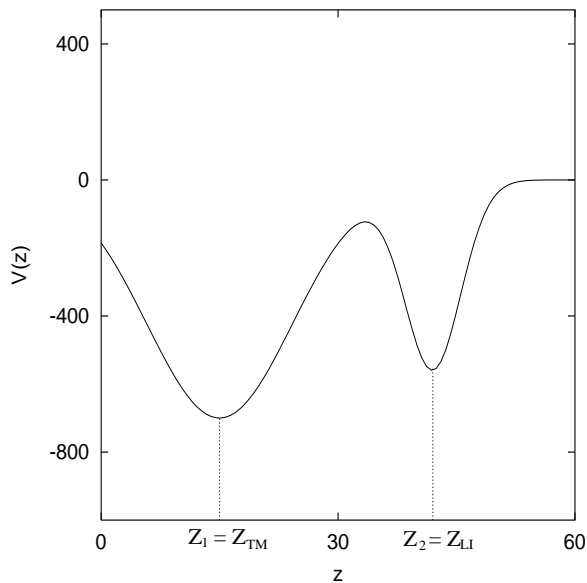}
\caption{\label{fig:epsart} The effective potential $V(z)$ in Eq.~(\ref{V(z)}).}
\end{figure}

Eq.~(\ref{dyn2}) was solved numerically by a finite difference method. 
Initially, the upper membrane was assigned a parabolic shape in
the contact region, and the intermembrane distance was kept constant ($\sim
50nm$) outside the contact region. For the membrane parameters
we use experimentally determined values \cite{simson} of 
$\gamma = 700 k_BT/\mu m^{2}$ and $\kappa = 400 k_BT$, and for the 
bond elasticity we take $\sigma_{1} = 13nm$ and $\sigma_{2} = 5nm$ \cite{chakraborty1}. 
Simulations were carried out for various ratios of $C^*_1/C^*_2$,
and local concentrations of the TCR-MHCp and
LFA1-ICAM1 were obtained from the intermembrane separation using Eqs.~(\ref{Rconstant}).

{\it Zero Noise:}
In the absence of noise ($\zeta=0$), Eq.~(\ref{dyn2}) is a relaxational dynamics 
toward the minimum of the potential $V(z)$, subject to the imposed boundary conditions.
For $C^*_1>C^*_2$ the minimum at $z_{TM}$ is deeper, and indeed in the numerical simulations
a central cluster of TCR-MHCp complex forms quickly in
the intercellular contact area. As the boundary of the contact
region is kept at a constant height close to $z_{LI}$, the longer 
LFA1-ICAM1 moves toward the boundary giving rise to a structure very similar to the final 
synaptic pattern obtained 
in the experimental and numerical studies \cite{grakoui,chakraborty1,chakraborty2,weikl}.

When the concentration $C^*_1$ is below $C^*_2$, the global minimum of
$V(z)$ switches to $z_{LI}$. In this regime, 
the receptor-ligand complex LFA1-ICAM1 concentration dominates the entire 
intercellular junction, and no synaptic pattern is formed. 
In the noise-less (mean-field) limit the location of the transition between the
two patterns is always at $C^*_1/C^*_2=1$. 
The patterns obtained and the conditions for transitions between them are 
quite different when $\zeta$ is finite.

{\it Finite Noise:}
In this case, numerical simulations of Eq.~(\ref{dyn2}) are performed with a
random $\zeta$ with zero mean, and variance equal to $k_BT$ to mimic thermal noise.
Starting with sufficiently large values of $C^*_1\geq C^*_2$, we again find that 
TCR-MHCp aggregates at the center of the intercellular junction at long times.
In Fig.~(2a) a cross section of the upper membrane and the associated protein
concentrations are shown. Very similar synaptic patterns have been observed 
\cite{monks,grakoui,krummel1,lee,stoll} for
{\it in vivo} and {\it in vitro} experiments with mature T cells.

\begin{figure}[h]
\includegraphics[height=3in,width=3in,angle=0]{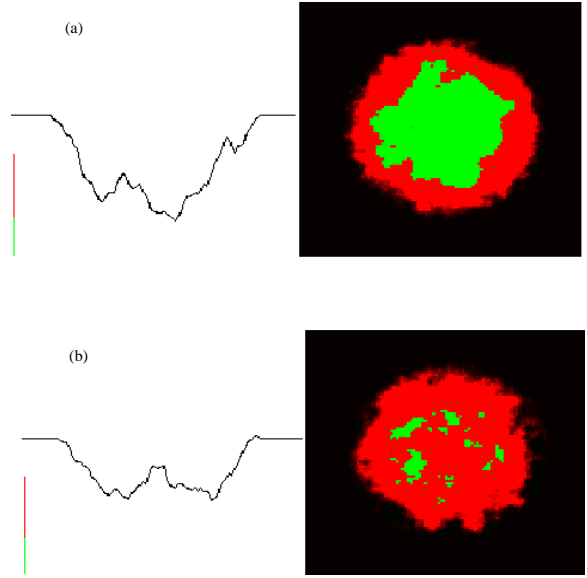}
\caption{\label{fig:epsart} Right panel: (a) Structure of a mature synapse: TCR-MHCp occupies
the central region (represented in green), surrounded by a ring of LFA1-ICAM1
(shown in red). (b) Dynamic accumulation of TCR-MHCp (green) at random
places in a sea of LFA1-ICAM1 (red) molecules as observed for thymocytes. 
Left panel: Height fluctuations corresponding to (a) and (b). The green 
($z_{TM}=15nm$) and red ($z_{LI}=42nm$) lines are meant to guide the eye.}
\end{figure}

For double positive (DP) thymocytes, the TCR concentration is much lower
compared to mature T-cells \cite{hailman}. This experimental fact corresponds to
lower values of $C^*_1$. When we numerically solve 
Eq.~(\ref{dyn2}) with $C^*_1/C^*_2<f_c$, 
the mature synapse is never formed. 
Rather, we observe (see Fig.~2b) fluctuating patterns with fleeting clusters of 
TCR-MHCp forming at various locations in a sea of LFA1-ICAM1. 
Note that this pattern is different in character from the one observed for 
$C^*_1/C^*_2<1$ in the noiseless case. Such dynamic patterns have been observed in 
recent experiments using DP thymocytes \cite{hailman,richie}, 
and can be understood using 
the simple picture of relaxational motion in a potential with two minima. 
As we lower the TCR concentration, the minimum at $z=z_{TM}$ becomes shallower,
and eventually the minimum near $z=z_{LI}$ corresponds to the global equilibrium.
However, thermal fluctuations allow sampling of the region near the 
minimum at $z=z_{TM}$. This will
favor sporadic aggregation of TCR-MHCp at random places. As
long as the barrier between the two minima is not very large, such
transient fluctuations will be frequent. 

The effects of thermal fluctuations can be estimated by 
expanding $V(z)$ to quadratic order around each of its minima.
The total energy cost 
of a deformation of wavenumber $p$ is given by
$E_i(p)=\left(k_BTC^*_i\sigma_i^{-2}+\gamma p^2+\kappa p^4\right)|z(p)|^2$.
Integrating out these (Gaussian) fluctuations yields a free energy per unit area of
\beq\label{Fi}
\frac{F^*_i}{k_BT}=C^*_i-\frac{1}{2}\int\frac{dp~p}{2\pi}\ln\left(
\frac{k_BTC^*_i\sigma_i^{-2}+\gamma p^2+\kappa p^4}{2\pi k_BT}\right).
\eeq
The location of the transition between the two classes of patterns can now be estimated 
from $F^*_1=F^*_2$. 
Thermal noise enhances the minimum with larger $\sigma$ (width),
because of its larger entropy. Hence the more flexible bond is formed more 
easily when thermal noise is present \cite{chen}. Fig.~3 shows that the larger the temperature, 
the smaller the value of $f_{c}$.  
This ``phase diagram'' could serve as a guide for the design of synthetic systems as it shows how 
to manipulate conditions such that synaptic patterns characteristic of mature T cells or thymocytes 
are realized.

\begin{figure}[h]
\includegraphics[height=3in,width=3in,angle=0]{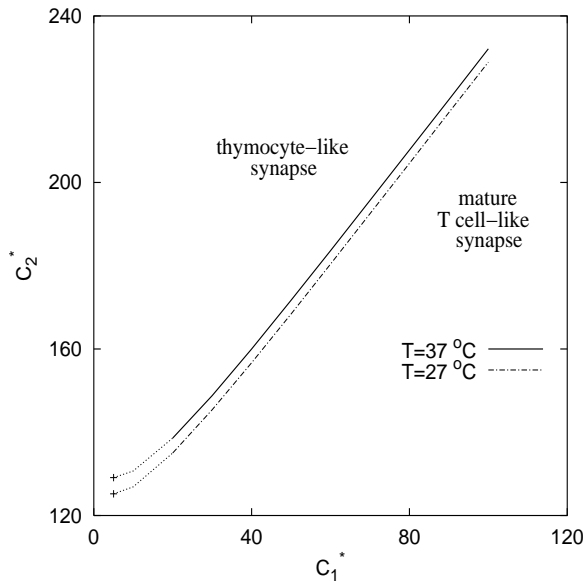}
\caption{\label{fig:epsart} Phase diagram (projection on the $C_{1}^*$-$C_{2}^*$ plane) for 
physiological temperature $T=37^{o} C$ and room temperature $T=27^{o} C$. $C_{1}^*$ and $C_{2}^*$
are measured in units of $(\mu m)^{-2}$. The dotted portion of the
curves (for low values of $C_{1}^*$-$C_{2}^*$) is conjectured as explained in the text.}
\end{figure}

Self consistency of the above quadratic expansion requires that the extent of 
the fluctuations should be less than the corresponding variance of the bond.
The mean square width of the interface trapped
around $z=z_i$ is given by
\beq\label{Wi}
W_i^2=\int\frac{dp~p}{2\pi}
\frac{k_BT}{k_BTC^*_i\sigma_i^{-2}+\gamma p^2+\kappa p^4}.
\eeq
Typical values of $C^*_1$ range from $O(10)(\mu m)^{-2}$ for thymocytes to $O(100)(\mu m)^{-2}$
in mature T cells. For this range of parameters, the constant term in the denominator
of Eq.~(\ref{Wi}) dominates, until it is cut-off by the bending rigidity at distances of
the order of $1\mu m$. (The overall size of the synapse is of the order of $10\mu m$ \cite{grakoui}.)
It is then easy to estimate the value of the integral, and we find that the ratios
$W_1/\sigma_1$ and $W_2/\sigma_2$ are of the order of 0.1.
While these estimated fluctuations are smaller than the width of the attractive potentials,
they are significant enough to make nonlinear corrections to the above quadratic expansion 
important. Thus the phase diagram in Fig.~3, using
Eq.~(\ref{Fi}) should be regarded as an approximation.

When anharmonic corrections are significant,
an alternative approach is to employ a functional renormalization group (RG) scheme 
used \cite{dfisher,lipowski1,lipowski2} in the context of wetting transitions.
It  consists of the following steps: 
{\em (1)} The fluctuating field $z(x)$ is divided into two parts 
corresponding to small wavenumbers ($0 < |p| < \Lambda/b$) 
and large wavenumbers ($\Lambda/b < |p| < \Lambda$); 
{\em (2)} The large wavenumber fluctuations $\Lambda/b < |p| < \Lambda$  
are integrated out to yield a coarse grained Hamiltonian; and 
{\em (3)} The system is rescaled to new coordinates $x'=x/b$ and $z'(x') = z(x)$. 
(The field $z$ does not need to be scaled in two dimensions.)
The second step can only be done approximately, and a first order expansion
of the potential $V(z)$ leads to the linear approximation\cite{dfisher}
\beq\label{RGV}
V_b(z)\approx{\cal L}[V(z)] = b^{2} \int_{- \infty}^{\infty} \frac{dz'}{\sqrt {2 \pi} \delta(b)}
        \exp\left[-\frac{ ( z - z')^{2} }{2 \delta^{2}(b)}\right]V(z').
\eeq
The width of the convolution, $\delta(b)$, is
\beq
\delta^{2} (b) = \int_{\Lambda/b}^\Lambda\frac{dp~p}{2\pi}
\frac{k_BT}{\gamma p^2+\kappa p^4}=
\frac{k_BT}{4\pi\gamma}\left[2\ln b-\ln\left(
\frac{1+\kappa \Lambda^2/\gamma}{1+\kappa \Lambda^2/(b^2\gamma)}
\right)\right].
\eeq
Note that $\kappa$ is an irrelevant operator in the RG sense (vanishing as $b^{-2}$
under scaling).
Indeed the second term in the above equation can be ignored asymptotically for
$b\gg 1$, but may be important for quantitative comparisons when the bending rigidity is large.

Applying Eq.~(\ref{RGV}) to the potential in Eq.~(\ref{V(z)}) yields the renormalized potential
\begin{widetext}
\beq\label{VR}
{V_b(z)} \approx b^2k_BT\left[ -\frac{\sigma_{1} C^*_{1}}
   { \sqrt{\sigma_{1}^{2} + \delta^{2}}}
e^{-\frac{(z-z_{1})^2}{2 (\sigma_{1}^{2}+\delta^{2})}}
 - \frac{ \sigma_{2} C^*_{2}}
   {\sqrt{\sigma_{2}^{2} + \delta^{2}}}
e^{ -\frac{(z-z_{2})^2}{2 (\sigma_{2}^{2}+\delta^{2})} }
 + c  \,{\rm erf}\left(z,\delta\right)\right].
\eeq
\end{widetext}
The soft wall renormalizes into an error function of width $\delta$, while
the  variance of the Gaussian potential is increased by $\delta^2$.
Thus, the bigger the width of the potential, the deeper it gets
due to the fluctuations, as it allows more flexibility (entropy) in bond formation.
For large values of the width $\delta$ (highly fluctuating surfaces), the location
of the transition point is shifted to $C^*_1/C^*_2=f_c=\sigma_2/\sigma_1\approx 0.38$.
Due to the factor of $b^2$ in Eq.~(\ref{VR}) the net potential gets stronger under RG,
and the linear approximation eventually breaks down. 

The RG transformation should be extended to a rescaling factor $b$ at which
the curvature at the minimum of the renormalized potential is such that the resulting
fluctuations in width are comparatively small.
Beyond this scale,  a  quadratic expansion around the minimum of the effective potential
should be valid.
As demonstrated previously, typical parameters for the synapse yield a bare potential
at which the above criterion is (barely) satisfied. 
Thus the linear RG approximation is of limited practical value in this context.
It may, however, yield interesting information about the global form of the phase
diagram as the concentrations $C^*_1$ and $C^*_2$ are further reduced.
From the double well shape of the potential, we anticipate that the overall phase 
diagram consists of a line of discontinuous transitions separating preferences in
the synapse for TM or LI bonds. This line is located at $C^*_1/C^*_2\leq 1$ for
large concentrations, and moves to   $C^*_1/C^*_2\approx \sigma_1/\sigma_2$ at
lower concentrations. It presumably terminates at an Ising critical point when the
fluctuations in width become of the order of the separation $z_{TM}-z_{LI}$ between
the two minima. Sufficiently small values of $C^*_1$ and $C^*_2$ should also result in
an unbinding transition which is located at 
$C^*_1\sigma_1+C^*_2\sigma_2\propto\sqrt{k_BT/\gamma}$.
It is thus useful to make a more detailed study of the phase diagram of the model
synapse as well as the dynamic trajectories as various phase boundaries are approached.

\bibliography{immunesynapse}    

\begin{thebibliography}{20}
\expandafter\ifx\csname natexlab\endcsname\relax\def\natexlab#1{#1}\fi
\expandafter\ifx\csname bibnamefont\endcsname\relax
  \def\bibnamefont#1{#1}\fi
\expandafter\ifx\csname bibfnamefont\endcsname\relax
  \def\bibfnamefont#1{#1}\fi
\expandafter\ifx\csname citenamefont\endcsname\relax
  \def\citenamefont#1{#1}\fi
\expandafter\ifx\csname url\endcsname\relax
  \def\url#1{\texttt{#1}}\fi
\expandafter\ifx\csname urlprefix\endcsname\relax\def\urlprefix{URL }\fi
\providecommand{\bibinfo}[2]{#2}
\providecommand{\eprint}[2][]{\url{#2}}

\bibitem[{\citenamefont{Abbas et~al.}(2000)\citenamefont{Abbas, Lictman, and
  Prober}}]{abbas}
\bibinfo{author}{\bibfnamefont{A.~K.} \bibnamefont{Abbas}},
  \bibinfo{author}{\bibfnamefont{A.~H.} \bibnamefont{Lictman}},
  \bibnamefont{and} \bibinfo{author}{\bibfnamefont{J.~S.}
  \bibnamefont{Prober}}, \emph{\bibinfo{title}{Cellular and Molecular
  Immunology}} (\bibinfo{publisher}{W.B. Saunders},
  \bibinfo{address}{Philadelphia}, \bibinfo{year}{2000}).

\bibitem[{\citenamefont{Monks et~al.}(1998)\citenamefont{Monks, Frieberg,
  Kupfer, Sciaky, and Kupfer}}]{monks}
\bibinfo{author}{\bibfnamefont{C.~R.~F.} \bibnamefont{Monks}},
  \bibinfo{author}{\bibfnamefont{B.~A.} \bibnamefont{Frieberg}},
  \bibinfo{author}{\bibfnamefont{H.}~\bibnamefont{Kupfer}},
  \bibinfo{author}{\bibfnamefont{N.}~\bibnamefont{Sciaky}}, \bibnamefont{and}
  \bibinfo{author}{\bibfnamefont{A.}~\bibnamefont{Kupfer}},
  \bibinfo{journal}{Nature} \textbf{\bibinfo{volume}{395}}, \bibinfo{pages}{82}
  (\bibinfo{year}{1998}).

\bibitem[{\citenamefont{Grakoui et~al.}(1999)\citenamefont{Grakoui, Bromley,
  Saumen, Davis, Shaw, Allen, and Dustin}}]{grakoui}
\bibinfo{author}{\bibfnamefont{A.}~\bibnamefont{Grakoui}},
  \bibinfo{author}{\bibfnamefont{S.~K.} \bibnamefont{Bromley}},
  \bibinfo{author}{\bibfnamefont{C.}~\bibnamefont{Saumen}},
  \bibinfo{author}{\bibfnamefont{M.~M.} \bibnamefont{Davis}},
  \bibinfo{author}{\bibfnamefont{A.~S.} \bibnamefont{Shaw}},
  \bibinfo{author}{\bibfnamefont{P.~M.} \bibnamefont{Allen}}, \bibnamefont{and}
  \bibinfo{author}{\bibfnamefont{M.~L.} \bibnamefont{Dustin}},
  \bibinfo{journal}{Science} \textbf{\bibinfo{volume}{285}},
  \bibinfo{pages}{221} (\bibinfo{year}{1999}).

\bibitem[{\citenamefont{Krummel et~al.}(2000)\citenamefont{Krummel, Sjaastad,
  Wulfing, and Davis}}]{krummel1}
\bibinfo{author}{\bibfnamefont{M.~F.} \bibnamefont{Krummel}},
  \bibinfo{author}{\bibfnamefont{M.~D.} \bibnamefont{Sjaastad}},
  \bibinfo{author}{\bibfnamefont{C.}~\bibnamefont{Wulfing}}, \bibnamefont{and}
  \bibinfo{author}{\bibfnamefont{M.~M.} \bibnamefont{Davis}},
  \bibinfo{journal}{Science} \textbf{\bibinfo{volume}{289}},
  \bibinfo{pages}{1349} (\bibinfo{year}{2000}).

\bibitem[{\citenamefont{Lee et~al.}(2002)\citenamefont{Lee, Holdorf, Dustin,
  Chan, Allen, and Shaw}}]{lee}
\bibinfo{author}{\bibfnamefont{K.~H.} \bibnamefont{Lee}},
  \bibinfo{author}{\bibfnamefont{A.~D.} \bibnamefont{Holdorf}},
  \bibinfo{author}{\bibfnamefont{M.~L.} \bibnamefont{Dustin}},
  \bibinfo{author}{\bibfnamefont{A.~C.} \bibnamefont{Chan}},
  \bibinfo{author}{\bibfnamefont{P.~M.} \bibnamefont{Allen}}, \bibnamefont{and}
  \bibinfo{author}{\bibfnamefont{A.~S.} \bibnamefont{Shaw}},
  \bibinfo{journal}{Science} \textbf{\bibinfo{volume}{295}},
  \bibinfo{pages}{1539} (\bibinfo{year}{2002}).

\bibitem[{\citenamefont{Stoll et~al.}(2001)\citenamefont{Stoll, Delon, Brotz,
  and Germain}}]{stoll}
\bibinfo{author}{\bibfnamefont{S.}~\bibnamefont{Stoll}},
  \bibinfo{author}{\bibfnamefont{J.}~\bibnamefont{Delon}},
  \bibinfo{author}{\bibfnamefont{T.~M.} \bibnamefont{Brotz}}, \bibnamefont{and}
  \bibinfo{author}{\bibfnamefont{R.~N.} \bibnamefont{Germain}},
  \bibinfo{journal}{Immunity} \textbf{\bibinfo{volume}{15}},
  \bibinfo{pages}{691} (\bibinfo{year}{2001}).

\bibitem[{\citenamefont{Hailman et~al.}(2002)\citenamefont{Hailman, Burack,
  Shaw, Dustin, and Allen}}]{hailman}
\bibinfo{author}{\bibfnamefont{E.}~\bibnamefont{Hailman}},
  \bibinfo{author}{\bibfnamefont{W.~R.} \bibnamefont{Burack}},
  \bibinfo{author}{\bibfnamefont{A.~S.} \bibnamefont{Shaw}},
  \bibinfo{author}{\bibfnamefont{M.~L.} \bibnamefont{Dustin}},
  \bibnamefont{and} \bibinfo{author}{\bibfnamefont{P.~M.} \bibnamefont{Allen}},
  \bibinfo{journal}{Immunity} \textbf{\bibinfo{volume}{16}}, \bibinfo{pages}{1}
  (\bibinfo{year}{2002}).

\bibitem[{\citenamefont{Ritchie et~al.}(2002)\citenamefont{Ritchie, Ebert, Wu,
  Krummel, Owen, and Davis}}]{richie}
\bibinfo{author}{\bibfnamefont{L.~I.} \bibnamefont{Ritchie}},
  \bibinfo{author}{\bibfnamefont{P.~J.~R.} \bibnamefont{Ebert}},
  \bibinfo{author}{\bibfnamefont{L.~C.} \bibnamefont{Wu}},
  \bibinfo{author}{\bibfnamefont{M.~F.} \bibnamefont{Krummel}},
  \bibinfo{author}{\bibfnamefont{J.~T.} \bibnamefont{Owen}}, \bibnamefont{and}
  \bibinfo{author}{\bibfnamefont{M.~M.} \bibnamefont{Davis}},
  \bibinfo{journal}{Immunity} \textbf{\bibinfo{volume}{16}},
  \bibinfo{pages}{595} (\bibinfo{year}{2002}).

\bibitem[{\citenamefont{Qi et~al.}(2001)\citenamefont{Qi, Groves, and
  Chakraborty}}]{chakraborty1}
\bibinfo{author}{\bibfnamefont{S.}~\bibnamefont{Qi}},
  \bibinfo{author}{\bibfnamefont{J.~T.} \bibnamefont{Groves}},
  \bibnamefont{and} \bibinfo{author}{\bibfnamefont{A.~K.}
  \bibnamefont{Chakraborty}}, \bibinfo{journal}{Proc. Natl. Acad. Sci. (USA)}
  \textbf{\bibinfo{volume}{98}}, \bibinfo{pages}{6548} (\bibinfo{year}{2001}).

\bibitem[{\citenamefont{Wulfing and Davis}(1998)}]{wulfing1}
\bibinfo{author}{\bibfnamefont{C.}~\bibnamefont{Wulfing}} \bibnamefont{and}
  \bibinfo{author}{\bibfnamefont{M.~M.} \bibnamefont{Davis}},
  \bibinfo{journal}{Science} \textbf{\bibinfo{volume}{283}},
  \bibinfo{pages}{2266} (\bibinfo{year}{1998}).

\bibitem[{\citenamefont{Dustin and Shaw}(1999)}]{shaw}
\bibinfo{author}{\bibfnamefont{M.~L.} \bibnamefont{Dustin}} \bibnamefont{and}
  \bibinfo{author}{\bibfnamefont{A.~S.} \bibnamefont{Shaw}},
  \bibinfo{journal}{Science} \textbf{\bibinfo{volume}{283}},
  \bibinfo{pages}{649} (\bibinfo{year}{1999}).

\bibitem[{\citenamefont{Hori et~al.}(2002)\citenamefont{Hori, Raychaudhuri, and
  Chakraborty}}]{chakraborty4}
\bibinfo{author}{\bibfnamefont{Y.}~\bibnamefont{Hori}},
  \bibinfo{author}{\bibfnamefont{S.}~\bibnamefont{Raychaudhuri}},
  \bibnamefont{and} \bibinfo{author}{\bibfnamefont{A.~K.}
  \bibnamefont{Chakraborty}}, \bibinfo{journal}{J. Chem. Phys.}
  \textbf{\bibinfo{volume}{117}}, \bibinfo{pages}{9491} (\bibinfo{year}{2002}).

\bibitem[{\citenamefont{Garboczi et~al.}(1996)\citenamefont{Garboczi, Ghosh,
  Utz, Fan, Biddison, and Wiley}}]{garboczi}
\bibinfo{author}{\bibfnamefont{D.~N.} \bibnamefont{Garboczi}},
  \bibinfo{author}{\bibfnamefont{P.}~\bibnamefont{Ghosh}},
  \bibinfo{author}{\bibfnamefont{U.}~\bibnamefont{Utz}},
  \bibinfo{author}{\bibfnamefont{Q.~R.} \bibnamefont{Fan}},
  \bibinfo{author}{\bibfnamefont{W.~E.} \bibnamefont{Biddison}},
  \bibnamefont{and} \bibinfo{author}{\bibfnamefont{D.~C.} \bibnamefont{Wiley}},
  \bibinfo{journal}{Nature} \textbf{\bibinfo{volume}{384}},
  \bibinfo{pages}{134} (\bibinfo{year}{1996}).

\bibitem[{\citenamefont{Simson et~al.}(1998)\citenamefont{Simson, Wallraff,
  Faix, Niewohner, Gerisch, and Sackmann}}]{simson}
\bibinfo{author}{\bibfnamefont{R.}~\bibnamefont{Simson}},
  \bibinfo{author}{\bibfnamefont{E.}~\bibnamefont{Wallraff}},
  \bibinfo{author}{\bibfnamefont{J.}~\bibnamefont{Faix}},
  \bibinfo{author}{\bibfnamefont{J.}~\bibnamefont{Niewohner}},
  \bibinfo{author}{\bibfnamefont{G.}~\bibnamefont{Gerisch}}, \bibnamefont{and}
  \bibinfo{author}{\bibfnamefont{E.}~\bibnamefont{Sackmann}},
  \bibinfo{journal}{Biphys. J.} \textbf{\bibinfo{volume}{74}},
  \bibinfo{pages}{514} (\bibinfo{year}{1998}).

\bibitem[{\citenamefont{Chakraborty}(2002)}]{chakraborty2}
\bibinfo{author}{\bibfnamefont{A.~K.} \bibnamefont{Chakraborty}},
  \bibinfo{journal}{Science STKE} \textbf{\bibinfo{volume}{2002}},
  \bibinfo{pages}{PE10} (\bibinfo{year}{2002}).

\bibitem[{\citenamefont{Weikl et~al.}(2002)\citenamefont{Weikl, Groves, and
  Lipowsky}}]{weikl}
\bibinfo{author}{\bibfnamefont{T.~R.} \bibnamefont{Weikl}},
  \bibinfo{author}{\bibfnamefont{J.~T.} \bibnamefont{Groves}},
  \bibnamefont{and} \bibinfo{author}{\bibfnamefont{R.}~\bibnamefont{Lipowsky}},
  \bibinfo{journal}{Europhys. Lett.} \textbf{\bibinfo{volume}{59}},
  \bibinfo{pages}{916} (\bibinfo{year}{2002}).

\bibitem[{\citenamefont{Chen}(2002)}]{chen}
\bibinfo{author}{\bibfnamefont{H.-Y.} \bibnamefont{Chen}},
  \bibinfo{journal}{cond-mat/0210398}  (\bibinfo{year}{2002}).

\bibitem[{\citenamefont{Fisher and Huse}(1985)}]{dfisher}
\bibinfo{author}{\bibfnamefont{D.~S.} \bibnamefont{Fisher}} \bibnamefont{and}
  \bibinfo{author}{\bibfnamefont{D.~A.} \bibnamefont{Huse}},
  \bibinfo{journal}{Phys. Rev. B} \textbf{\bibinfo{volume}{32}},
  \bibinfo{pages}{247} (\bibinfo{year}{1985}).

\bibitem[{\citenamefont{Lipowsky and Fisher}(1986)}]{lipowski1}
\bibinfo{author}{\bibfnamefont{R.}~\bibnamefont{Lipowsky}} \bibnamefont{and}
  \bibinfo{author}{\bibfnamefont{M.~E.} \bibnamefont{Fisher}},
  \bibinfo{journal}{Phys. Rev. Lett.} \textbf{\bibinfo{volume}{10}},
  \bibinfo{pages}{2411} (\bibinfo{year}{1986}).

\bibitem[{\citenamefont{Lipowsky and Fisher}(1987)}]{lipowski2}
\bibinfo{author}{\bibfnamefont{R.}~\bibnamefont{Lipowsky}} \bibnamefont{and}
  \bibinfo{author}{\bibfnamefont{M.~E.} \bibnamefont{Fisher}},
  \bibinfo{journal}{Phys. Rev. B} \textbf{\bibinfo{volume}{36}},
  \bibinfo{pages}{2126} (\bibinfo{year}{1987}).

\end{thebibliography}

\end{document}